\documentclass[aps,prb,preprint,superscriptaddress]{revtex4}

\usepackage[dvips]{graphicx}

\begin{document}

\title{Statistical analysis of time-resolved emission from ensembles of semiconductor quantum dots: interpretation of exponential decay models}

\author{A. F. van Driel}
\affiliation{Debye Institute, Utrecht University, P.O. Box 80 000,
3508 TA Utrecht, The Netherlands}

\author{I. S. Nikolaev}
\affiliation{Center for Nanophotonics, FOM Institute for Atomic
and Molecular Physics (AMOLF), 1098 SJ Amsterdam, The
Netherlands}\affiliation{Complex Photonic Systems (COPS),
Department of Science and Technology and MESA+ Research Institute,
University of Twente, 7500 AE Enschede, The Netherlands}
\author{P. Vergeer}
\affiliation{Debye Institute, Utrecht University, P.O. Box 80 000,
3508 TA Utrecht, The Netherlands}
\author{P. Lodahl}
\affiliation{Complex Photonic Systems (COPS), Department of
Science and Technology and MESA+ Research Institute, University of
Twente, 7500 AE Enschede, The
Netherlands}\affiliation{COM$\cdot$DTU Department of
Communications, Optics, and Materials, Nano$\cdot$DTU, Technical
University of Denmark, Denmark}
\author{D. Vanmaekelbergh}
\affiliation{Debye Institute, Utrecht University, P.O. Box 80 000,
3508 TA Utrecht, The Netherlands}
\author{W. L. Vos}\email[Electronic address: ]{w.l.vos@utwente.nl}
\homepage[Webpage: ]{www.photonicbandgaps.com}
\affiliation{Center for Nanophotonics, FOM Institute for Atomic
and Molecular Physics (AMOLF), 1098 SJ Amsterdam, The
Netherlands}\affiliation{Complex Photonic Systems (COPS),
Department of Science and Technology and MESA+ Research Institute,
University of Twente, 7500 AE Enschede, The Netherlands}

\date{Prepared for Phys. Rev. B in October, 2006.}

\begin{abstract}
We present a statistical analysis of time-resolved spontaneous
emission decay curves from ensembles of emitters, such as
semiconductor quantum dots, with the aim to interpret ubiquitous
non-single-exponential decay. Contrary to what is widely assumed,
the density of excited emitters and the intensity in an emission
decay curve are not proportional, but the density is a
time-integral of the intensity. The integral relation is crucial
to correctly interpret non-single-exponential decay. We derive the
proper normalization for both a discrete, and a continuous
distribution of rates, where every decay component is multiplied
with its radiative decay rate. A central result of our paper is
the derivation of the emission decay curve in case that both
radiative and non-radiative decays are independently distributed.
In this case, the well-known emission quantum efficiency can not
be expressed by a single number anymore, but it is also
distributed. We derive a practical description of
non-single-exponential emission decay curves in terms of a single
distribution of decay rates; the resulting distribution is
identified as the distribution of total decay rates weighted with
the radiative rates. We apply our analysis to recent examples of
colloidal quantum dot emission in suspensions and in photonic
crystals, and we find that this important class of emitters is
well described by a log-normal distribution of decay rates with a
narrow and a broad distribution, respectively. Finally, we briefly
discuss the Kohlrausch stretched-exponential model, and find that
its normalization is ill-defined for emitters with a realistic
quantum efficiency of less than 100 $\%$.
\end{abstract}


\maketitle

\section{Introduction}
Understanding the decay dynamics of excited states in emitters
such as semiconductor quantum dots is of key importance for
getting insight in many physical, chemical and biological
processes. For example, in biophysics the influence of F\"orster
resonance energy transfer on the decay dynamics of donor molecules
is studied to quantify molecular
dynamics\cite{Lakowicz,2005Medintz}. In cavity quantum
electrodynamics, modification of the density of optical modes
(DOS) is quantified by measuring the decay dynamics of light
sources. According to Fermi's 'Golden Rule' the radiative decay
rate is proportional to the DOS at the location of the
emitter\cite{Loudon}. Nanocrystalline quantum
dots\cite{2002Crooker,2002Zhang,2005Medintz},
atoms\cite{1970Drexhage,1997Amos} and dye
molecules\cite{2002Danz,2002Astilean} are used as light sources in
a wide variety of systems. Examples of such systems are many
different kinds of photonic materials, including metallic and
dielectric
mirrors\cite{1970Drexhage,1997Amos,2002Danz,2002Astilean,2002Zhang},
cavities\cite{2001Bayer}, metallic
films\cite{2005Biteen,2005Song}, two\cite{2005Fujita,2005Kress}-,
and three-dimensional\cite{2004Lodahl} photonic crystals.

\begin{figure}
\includegraphics[width=0.8\columnwidth]{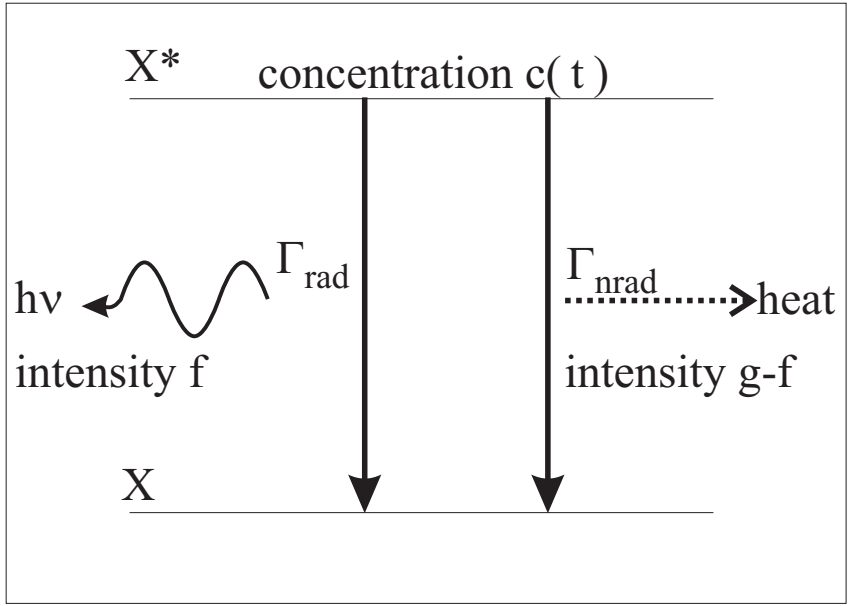}
\caption{\label{Scheme} Schematic of the relation between decay of
an excited state $X^{*}$ to the ground state $X$ and experimental
observable parameters. The density of emitters in the excited
state is equal to $c(t)$ which can be probed by transient
absorption. The emitted light intensity as a function of time
$f(t)$ is recorded in luminescence decay measurements. In
photothermal measurements the released heat $(g-f)(t)$ after
photoexcitation is detected. $g(t)$ describes the total decay,
i.e., the sum of the radiative and the non-radiative decay. }
\end{figure}

Figure \ref{Scheme} shows how observable parameters are related to
the decay of an excited state $X^*$ to the ground state $X$. In
photoluminescence lifetime measurements the decay of the number of
excited emitters is probed by recording a photoluminescence decay
curve ($f(t$)). The number of excited emitters $c(t)$ can be
probed directly by transient absorption
measurements\cite{1995Foggi,1998Klimov,2000Neuwahl} and
non-radiative decay ($g-f)(t)$ can be recorded with photothermal
techniques\cite{Rosencwaig,2003Grinberg} (see Fig. \ref{Scheme}).
$g(t)$ is here defined as the \emph{total} intensity, i.e., the
sum of the radiative and non-radiative processes. In this paper we
discuss photoluminescence lifetime measurements, which are
generally recorded by
time-correlated-single-photon-counting\cite{Lakowicz}. The decay
curve $f(t)$ consists of a histogram of the distribution of
arrival times of single photons after many excitation-detection
cycles\cite{Lakowicz}. The histogram is modelled with a
decay-function from which the decay time of the process is
deduced.

In the simplest case when the system is characterized by a single
decay rate $\Gamma$, the decay curve is described by a
single-exponential function. However, in many cases the decay is
much more complex and strongly differs from single-exponential
decay\cite{1986James,1990Siemiarczuk,1990Brochon,1995Foggi,2002Crooker,2003Wlodarczyk,2004Wuister,2004Lodahl}.
This usually means that the decay is characterized by a
distribution of rates instead of a single rate\footnote{In the
case of strong coupling in cavity quantum electrodynamics, the
decay of even a single emitter is not single-exponential.
Experimental situations where this may be encountered are emitters
in a high-finesse cavity, or van Hove singularities in the LDOS of
a photonic crystal.}. For example, ensembles of quantum dots in
photonic crystals experience the spatial and orientational
variations of the projected LDOS explaining the
non-single-exponential character of the decay\cite{2006Nikolaev}.
It is a general problem to describe such relaxation processes
which do not follow a simple single-exponential decay. Sometimes
double- and triple-exponential models are justified on the basis
of prior knowledge of the emitters\cite{Lakowicz}. However, in
many cases no particular multi-exponential model can be
anticipated on the basis of physical knowledge of the system
studied and a decision is made on basis of quality-of-fit.

Besides multi-exponential models, the stretched-exponential model
or Kohlrausch function\cite{1854Kohlrauscha} is frequently
applied. The stretched-exponential function has been applied to
model diffusion processes\cite{2001Deschenes}, dielectric
relaxation\cite{1980Lindsey}, capacitor
discharge\cite{1854Kohlrauschb}, optical Kerr effect
experiments\cite{2004Torre} and luminescence
decay\cite{2002Schlegel,2003Chen,2004Fisher}. The physical origin
of the apparent stretched-exponential decay in many processes
remains a source of intense
debate\cite{1985Huber,1991Alvarez,2002Lee}.

Surprisingly, in spite of the rich variety of examples where
non-single-exponential decay appears, there is no profound
analysis of the models available in the literature. Therefore, we
present in this paper a statistical analysis of time-resolved
spontaneous emission decay curves from ensembles of emitters with
the aim to interpret ubiquitous non-single-exponential decay.
Contrary to what is widely assumed, the density of excited
emitters $c(t)$ and the intensity in an emission decay curve
($f(t)$ or $g(t)$) are not proportional, but the density is a
time-integral of the intensity. The integral relation is crucial
to correctly interpret non-single-exponential decay. We derive the
proper normalization for both a discrete, and a continuous
distribution of rates, where every decay component is multiplied
with its radiative decay rate. A central result of our paper is
the derivation of the emission decay curve $f(t)$ in case that
both radiative and non-radiative decays are independently
distributed. In this most general case, the well-known emission
quantum efficiency is also distributed. Distributed radiative
decay is encountered in photonic media\cite{2006Nikolaev}, while
distributed non-radiative decay has been reported for colloidal
quantum dots\cite{2002Schlegel,2004Fisher} and powders doped with
rare-earths\cite{2005Vergeer}. We derive a practical description
of non-single-exponential emission decay curves in terms of the
distribution of total decay rates weighted with the radiative
rates. Analyzing decay curves in terms of distributions of decay
rates has the advantage that information on physically
interpretable rates is readily available, as opposed to previously
reported analysis in terms of lifetimes. We apply our analysis to
recent examples of colloidal quantum dot emission in suspensions
and in photonic crystals. We find excellent agreement with a
log-normal distribution of decay rates for such quantum dots. In
the final Section, we discuss the Kohlrausch stretched-exponential
model, and find that its normalization is ill-defined for emitters
with a realistic quantum efficiency of less than 100 $\%$.

\section{Decay models}
\subsection{Relation between the concentration of emitters and the decay curve}
\begin{figure}
\includegraphics[width=0.8\columnwidth]{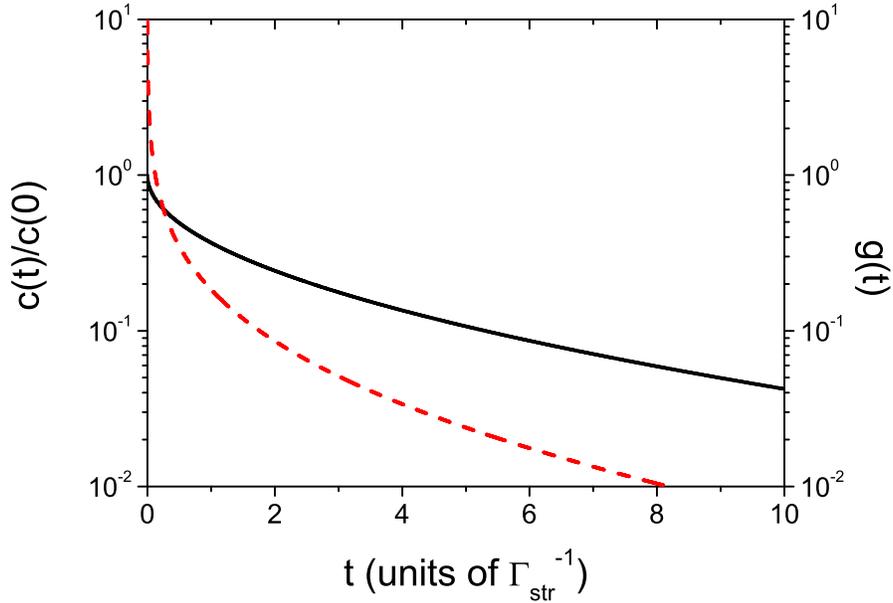}
\caption{\label{StrExpFg}(color online) Plot of a
non-single-exponential decay of the fraction $\frac{c(t)}{c(0)}$
(black solid curve, left axis) and the corresponding total
intensity decay curve $g(t)$ (red dashed curve, right axis). The
curves that describe the fraction of excited emitters and the
corresponding intensity decay curve are strongly different. In
this example, $\frac{c(t)}{c(0)}$ is the Kohlrausch
stretched-exponential decay of the fraction (Eq. \ref{str1}, black
solid curve) and $g(t)$ the corresponding decay curve  (Eq.
\ref{str2}, red dashed curve). We have taken $\beta$=0.5 and
$\Gamma_{str}=1$.}
\end{figure}

A decay curve is the probability density of emission which is
therefore modelled with a so-called probability density
function\cite{Dougherty}. This function tends to zero in the limit
$t\rightarrow\infty$. The decay of the fraction of excited
emitters $\frac{c(t^{'})}{c(0)}$ at time $t'$ is described with a
reliability function or cumulative distribution function
$\left(1-\frac{c(t^{'})}{c(0)}\right)$\cite{Dougherty}. Here
$c(0)$ is the concentration of excited emitters at $t'=0$. The
reliability function tends to one in the limit $t^{'}\rightarrow
\infty$ and to zero in the limit $t^{'}\rightarrow 0$. The
fraction of excited emitters and the decay curve, i.e., the
reliability function and the probability density
function\cite{Dougherty}, are related as follows:
\begin{equation}
\label{master}\int_{0}^{t^{'}} g(t)dt=1-\frac{c(t^{'})}{c(0)}
\end{equation}
Physically this equation means that the decrease of the
concentration of excited emitters at time $t^{'}$ is equal to the
integral of all previous decay events. Or equivalently: the total
intensity $g(t)$ is proportional to the time-derivative of the
fraction of excited emitters. As an illustration, Fig.
\ref{StrExpFg} shows a non-single-exponential decay function
simultaneously with the corresponding decay curve. It is clear
that both curves are strongly different. In many reports, however,
the distinction between the reliability function and the
probability density function is neglected: the intensity of the
decay curve $g(t)$ is taken to be directly proportional to the
fraction of excited emitters $\frac{c(t^{'})}{c(0)}$. This
proportionality only holds for single-exponential decay and not
for non-single-exponential decay, which has important consequences
for the interpretation of non-single-exponential decay.

\subsection{Single-exponential decay}
In this section, we will illustrate some concepts with the
well-known single-exponential model. We will also indicate which
features of single-exponential decay will break down in the
general case of non-single-exponential decay. It is well known
that in case of first-order kinetics the rate of decrease of the
concentration is constant in time:
\begin{equation}
\frac{d\,c(t^{'})}{dt^{'}}=-\Gamma c(t^{'})
\end{equation}
where $\Gamma$ is the decay rate of the process. As a consequence,
the concentration $c(t^{'})$ decreases single-exponentially in
time:
\begin{equation}\label{fracsinexp}
\frac{c(t^{'})}{c(0)}=\exp(-\Gamma t^{'})
\end{equation}
\begin{figure}
\includegraphics[width=0.8\columnwidth]{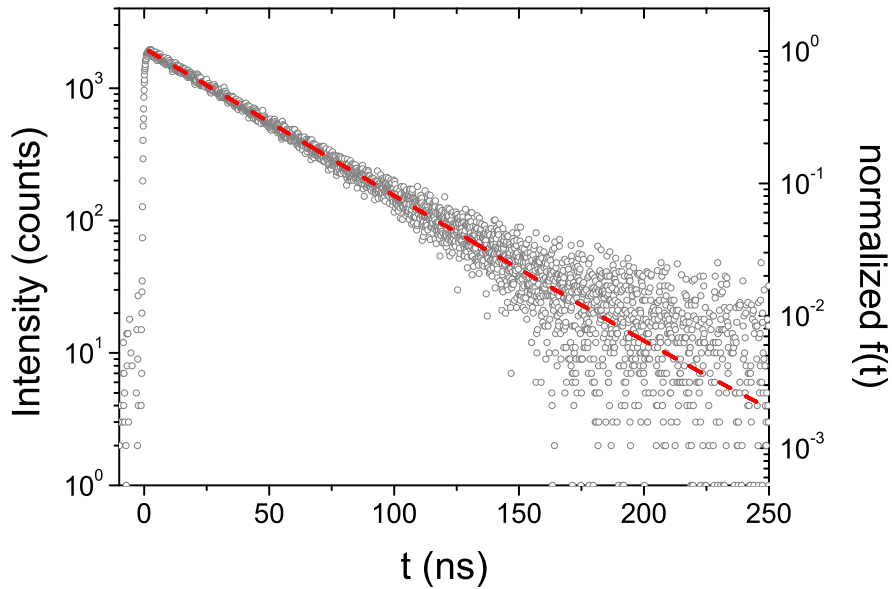}
\caption{\label{SinExp}(color online) Luminescence decay curve of
emission from a dilute suspension of CdSe quantum dots (open dots,
left axis). Data were collected at the red side of the emission
maximum of the suspension, at $\lambda$ = 650 $\pm$ 5 $nm$.
Single-exponential modelling (red dashed curve, right axis) yields
a decay time of 39.0$\pm2.8ns$ and a $\chi_{r}^{2}$ of 1.12. The
average photon arrival time $<t>$, calculated with Eq.
\ref{sinexp2}, is 39.1 $ns$.}
\end{figure}
The mathematical expression for the luminescence decay curve can
be obtained by inserting Eq. \ref{fracsinexp} into Eq.
\ref{master}, where $\Gamma$ is identified with the total decay
rate $\Gamma_{tot}$, resulting in:
\begin{equation}
\label{master2} g(t)=\Gamma_{rad}\exp(-\Gamma_{tot}
t)+\Gamma_{nrad}\exp(-\Gamma_{tot} t)
\end{equation}
where $\Gamma_{rad}$ is the radiative decay rate, $\Gamma_{nrad}$
is the nonradiative decay rate and $\Gamma_{tot}$ is the total
decay rate with $\Gamma_{tot}$ = $\Gamma_{rad}$ + $\Gamma_{nrad}$.
In a luminescence decay measurement the recorded signal is
proportional to the first term of $g(t)$ only which is $f(t)$:
\begin{equation}
\label{sinexp1} f(t)=\alpha\Gamma_{rad}\exp(-\Gamma_{tot} t)
\end{equation}
and therefore a single-exponential luminescence decay process is
modelled with Eq. \ref{sinexp1}. The pre-exponential factor
$\alpha$ is usually taken as adjustable parameter, and it is
related to several experimental parameters, i.e., the number of
excitation-emission cycles in the experiment, the
photon-collection efficiency and the concentration of the emitter.
Henceforth $\alpha$ will be omitted in our analysis. A comparison
between Eqs. \ref{sinexp1} and \ref{fracsinexp} shows that in the
case of pure single-exponential decay neglect of the distinction
between the reliability function (Eq. \ref{fracsinexp}) and the
probability density function (Eq. \ref{sinexp1}) has no important
consequences, since both the fraction and the decay curve are
single-exponential. As Fig. \ref{StrExpFg} shows, this neglect
breaks down in the case of non-single-exponential decay.

Figure \ref{SinExp} shows a luminescence decay curve of a dilute
suspension of CdSe quantum dots in chloroform at a wavelength of
$\lambda$ = 650 $\pm$ 5 $nm$\cite{2005vanDriel}, with the number
of counts on the ordinate and the time on the abscissa. Clearly,
the data agree well with single-exponential decay as indicated by
the quality-of-fit $\chi_{r}^{2}$ of 1.12, close to the ideal
value of 1. This means that all individual quantum dots that emit
light in this particular wavelength-range do so with the same rate
of $\frac{1}{39.0}$ $ns^{-1}$. It appears that the rate of
emission strongly depends on the emission frequency and that it is
determined by the properties of the bulk semiconductor
crystal\cite{2005vanDriel}.

Since $f(t)$ as given by Eq. \ref{sinexp1} is a probability
density function, the probability of emission in a certain
time-interval can be deduced by integration. The total probability
for emission at all times between $t=0$ and $t\rightarrow\infty$
is given by
\begin{equation}
\label{norml} \int_{0}^\infty f(t)dt=\int_{0}^\infty \Gamma_{rad}
\exp(-\Gamma_{tot} t)dt=\frac{\Gamma_{rad}}{\Gamma_{tot}}
\end{equation}
which is equal to the luminescence quantum efficiency. The
luminescence quantum efficiency is defined as the probability of
emission after excitation\cite{Lakowicz}. The correct recovery of
this result in Eq. \ref{norml} shows that Eq. \ref{sinexp1} is
properly normalized.

The average arrival time of the emitted photons or the average
decay time can be calculated by taking the first moment of Eq.
\ref{sinexp1}:
\begin{equation}
\label{sinexp2} <t>=\tau_{av}=\frac{\int_{0}^\infty
f(t)tdt}{\int_{0}^\infty f(t)dt}= \frac{1}{\Gamma_{tot}}
\end{equation}
Only in the case of single-exponential decay the average decay
time $<t>$ is equal to the inverse of the total decay rate
$\Gamma_{tot}$. The average arrival time for the data in Fig.
\ref{SinExp} was $<t>=39.1$ ns, very close to the value of 39.0
$\pm2.8ns$ obtained from single-exponential modelling, which
further confirms the single-exponential character of the decay of
quantum dots in suspension.

\subsection{Discrete distribution of decay rates}
\begin{figure}
\includegraphics[width=0.8\columnwidth]{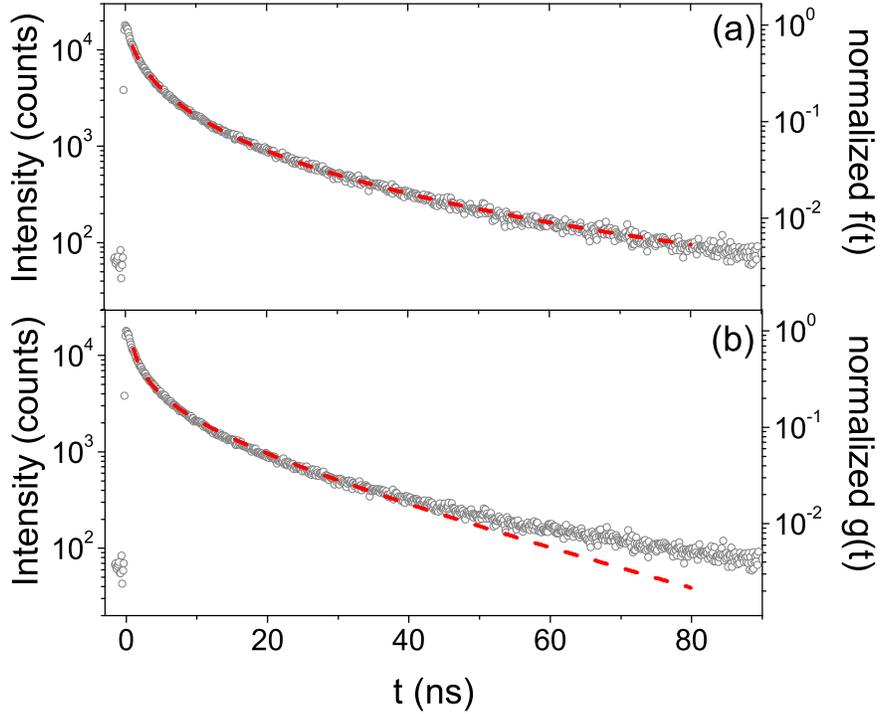}
\caption{\label{MultiExp}(color online) Luminescence decay curve
of emission from CdSe quantum dots in a titania inverse opal
photonic crystal (dots, left axis). The lattice parameter of the
titania inverse opal was 340 nm and the emission wavelength
$\lambda=595 nm$. (a) A log-normal distribution of rates (Eq.
\ref{lognor} and \ref{lognor2}, red dashed curve, right axis)
models the data extremely well ($\chi_{r}^{2}$=1.17). The
$\Gamma_{mf}$ is 91.7 $\mu s^{-1}$ ($\frac{1}{\Gamma_{mf}}=10.9$
$ns$) and the width of the distribution $\Delta\Gamma$ is 0.57
$ns^{-1}$. (b) In contrast, a Kohlrausch stretched-exponential
model (red dashed curve, right axis) does not fit the data
($\chi_{r}^2=60.7$). The stretched-exponential curve corresponds
to $\Gamma_{str}$=96.2 $\mu s^{-1}$ ($\frac{1}{\Gamma_{str}}$=10.4
ns), an average decay time $<t>$ of 31.1 $ns$, and a $\beta$-value
of 0.42.}
\end{figure}

In contrast to the example shown in Fig. \ref{SinExp}, there are
many cases in which decay curves cannot be modelled with a
single-exponential function. As an example, Fig. \ref{MultiExp}
shows a strongly non-single-exponential decay curve of spontaneous
emission from CdSe quantum dots in an inverse opal photonic
crystal\cite{2004Lodahl,2006Nikolaev}.
If a non-single-exponential decay curve is modelled with a sum of
single-exponentials, the decay curve has the following form:
\begin{equation}
\label{multi1}f(t)=\frac{1}{c(0)}\sum_{i=1}^{n}c_{i}\Gamma_{rad,i}\exp(-\Gamma_{tot,i}t)
\end{equation}
where $n$ is the number of different emitters (or alternatively
the number of different environments of single
emitters\cite{2006Nikolaev}), $c_{i}$ is the concentration of
emitters that has a radiative decay rate $\Gamma_{rad,i}$, and
$c(0)$ is the concentration of excited emitters at $t=0$, i.e.,
the sum of all concentrations $c_{i}$. When the different
fractions (or environments) are distributed in a particular way, a
distribution function $\rho(\Gamma_{tot})$ may be used. Such a
function describes the distribution or concentration of the
emitters over the emission decay rates at time $t=0$. The fraction
of emitters with a total decay rate $\Gamma_{tot,i}$ is equal to
\begin{eqnarray}
\label{distrib}\frac{c_{i}}{c(0)}&=&\frac{1}{c(0)}\frac{(c(\Gamma_{tot,i-1})+c(\Gamma_{tot,i+1}))}{2}
\\&=&\frac{1}{2}\int_{\Gamma_{tot,i-1}}^{\Gamma_{tot,i+1}}\rho(\Gamma_{tot})d\Gamma_{tot}\nonumber
\\ &=&\rho(\Gamma_{tot,i})\Delta\Gamma_{tot}\nonumber
\end{eqnarray}
where $\rho(\Gamma_{tot,i})$ expresses the distribution of the
various components $i$ over the rates $\Gamma_{tot,i}$ and has
units of inverse rate $s$. $\Delta\Gamma_{tot}$ is the separation
between the various components $i$ in the sum.
The decay curve now has the following mathematical form:
\begin{equation}
\label{multi}f(t)=\sum_{i=1}^{n}\Delta\Gamma_{tot}\rho(\Gamma_{tot,i})\Gamma_{rad,i}\exp(-\Gamma_{tot,i}t)
\end{equation}
It is important to note that in Eq. \ref{multi} every component in
the sum is correctly normalized since every component is
multiplied with its radiative decay rate $\Gamma_{rad,i}$.

\subsection{Continuous distribution of decay rates}
For infinitesimal values of $\Delta\Gamma_{tot}$, Eq. \ref{multi}
can be written as an integral:
\begin{equation}
\label{multiint}f(t)=\int_{0}^{\infty}\Gamma_{rad}(\Gamma_{tot})\rho(\Gamma_{tot})\exp(-\Gamma_{tot}t)d\Gamma_{tot}
\end{equation}
In the case of single-exponential decay the distribution function
is strongly peaked around a central $\Gamma_{tot}$-value, i.e.,
the distribution function is a Dirac delta function. Inserting a
Dirac delta function into Eq. \ref{multiint} recovers Eq.
\ref{sinexp1}:
\begin{eqnarray}
f(t)&=&\int_{0}^{\infty}\Gamma_{rad'}\,\delta(\Gamma_{tot}
-\Gamma_{tot'})\exp(-\Gamma_{tot}t)d\Gamma_{tot} \nonumber \\
\label{delta1}&=&\Gamma_{rad'}\exp(-\Gamma_{tot'}t)
\end{eqnarray}
This result confirms that the generalization to Eq. \ref{multiint}
is correct since it yields the correctly normalized
single-exponential functions.

In Eq. \ref{multiint} it is tacitly assumed that for every
$\Gamma_{tot}$ there is one $\Gamma_{rad}$: the function
$\Gamma_{rad}(\Gamma_{tot}$) relates each $\Gamma_{tot}$ to
exactly one $\Gamma_{rad}$. In general both $\Gamma_{tot}$ and
$\Gamma_{rad}$ vary independently, and Eq. \ref{multiint} is
generalized to
\begin{eqnarray}
&f(t)&\label{multiint3}\\
&=&\int_{0}^{\infty}\left[\int_{0}^{\Gamma_{tot}}d\Gamma_{rad}
\,\,\rho_{\Gamma_{tot}}(\Gamma_{rad})\,\,\Gamma_{rad}\right]\rho(\Gamma_{tot})\exp(-\Gamma_{tot}t)d\Gamma_{tot}
\nonumber
\end{eqnarray}
where $\rho_{\Gamma_{tot}}(\Gamma_{rad})$ is the normalized
distribution of $\Gamma_{rad}$ at constant $\Gamma_{tot}$. For
every $\Gamma_{tot}$ the integration is performed over all
radiative rates; a distribution of $\Gamma_{rad}$ is taken into
account for every $\Gamma_{tot}$. Eq. \ref{multiint3} is the most
general expression of a luminescence decay curve and a central
result of our paper. From this equation every decay curve with a
particular distribution of rates can be recovered. An example
described by Eq. \ref{multiint3} is an ensemble of quantum dots in
a photonic crystal. In photonic crystals the local density of
optical states (LDOS) varies with the location in the crystal and
the distribution of dipole orientations of the
emitters\cite{1996Sprik}. Therefore, an ensemble of emitters with
a certain frequency emit light with a distribution of radiative
rates $\Gamma_{rad}$. In addition, when an ensemble of emitters
has a distributed $\Gamma_{tot}$ and a single radiative rate
$\Gamma_{rad}$, i.e., $\rho_{\Gamma_{tot}}(\Gamma_{rad})$ is a
delta-function, then Eq.~\ref{multiint3} reduces to Eq.~
\ref{multiint}. Even though the non-radiative rates may still be
distributed, Eq.~\ref{multiint} suffices to describe the decay
curve since for every $\Gamma_{tot}$ there is only one
$\Gamma_{rad}$. Such a situation appears, for example, with
powders doped with rare earth ions\cite{2005Vergeer} and with
polymer films doped with quantum
dots\cite{2002Schlegel,2004Fisher}.

Interestingly, an ensemble of emitters with a distribution of
rates $\Gamma_{tot}$ is \emph{not} completely characterized by a
single value of the quantum efficiency (as opposed to
Eq.~\ref{norml} for single-exponential decay). In such an
ensemble, the quantum efficiency is distributed, since each
$\Gamma_{tot}$ is associated with a distribution of radiative
rates $\Gamma_{rad}$. The average quantum efficiency $<QE>$ can be
calculated by integrating Eq. \ref{multiint3} for all times:
\begin{eqnarray}
<QE>&=&\int_{0}^{\infty}f(t)dt\label{multiint5}\\
&=&\int_{0}^{\infty}\int_{0}^{\infty}\left[\int_{0}^{\Gamma_{tot}}d\Gamma_{rad}
\,\,\rho_{\Gamma_{tot}}(\Gamma_{rad})\,\,\Gamma_{rad}\right]\rho(\Gamma_{tot})\exp(-\Gamma_{tot}t)d\Gamma_{tot}dt
\nonumber
\end{eqnarray}

\begin{figure}
\includegraphics[width=0.8\columnwidth]{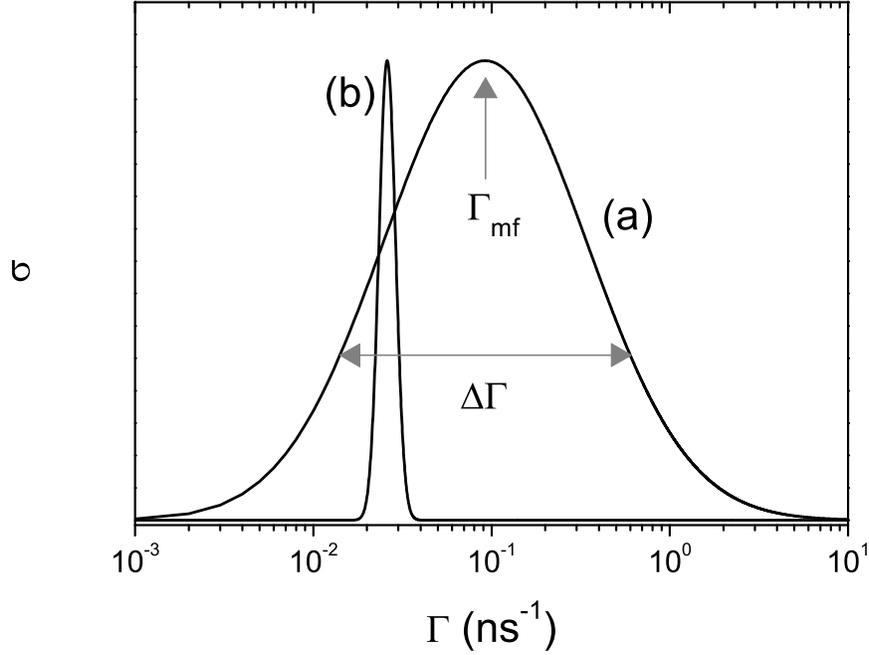}
\caption{\label{Distrib}Log-normal distribution of $\Gamma$. This
distribution was modelled to the data of Fig. \ref{MultiExp}
(curve a, quantum dots in photonic crystal) and Fig. \ref{SinExp}
(curve b, quantum dots in a diluted suspension), with
$\Gamma_{mf}$ and $\Delta\Gamma$ as adjustable parameters. For (a)
$\Gamma_{mf}$ is 91.7 $\mu s^{-1}$ ($\frac{1}{\Gamma_{mf}}=10.9$
$ns$) and the width of the distribution $\Delta\Gamma$ was 0.57
$ns^{-1}$ and for (b) $\Gamma_{mf}$ was 25.8 $\mu s^{-1}$
($\frac{1}{\Gamma_{mf}}=38.8$ $ns$) and the width of the
distribution $\Delta\Gamma$ was 0.079 $ns^{-1}$.}
\end{figure}

Most often, detailed information on the relation between
$\Gamma_{tot}$ and $\Gamma_{rad}$ is not available. Then,
modelling directly with a distribution of decay rates is
applied\cite{1985Jamesc,1986James,1990Brochon,2002Lee,2003Wlodarczyk}.
This approach has a major advantage over modelling with a
stretched-exponential function, where it is complicated to deduce
the distribution of decay rates (see below). A function of the
following form is used to model the non-single-exponential decay
curve:
\begin{equation}
\label{multi2}
f(t)=\int_{0}^{\infty}\sigma(\Gamma_{tot})\exp(-\Gamma_{tot}
t)d\Gamma_{tot}
\end{equation}
In Eq. \ref{multi2} the various components are not separately
normalized as in Eq. \ref{multiint3}. Modelling with Eq.
\ref{multi2} boils down to using an infinite series of
single-exponentials which are expressed with only a few free
parameters. The form of the distribution can usually not be
predicted and a decision is made on basis of quality-of-fit. While
a good fit does not prove that the chosen distribution is unique,
it does extract direct physical information from the
non-single-exponential decay on an ensemble of emitters and their
environment\cite{2006Nikolaev}.

It is widely assumed that $\sigma(\Gamma)$ is equal to the
distribution of total rates
\cite{1986James,1990Brochon,2001Lee,2003Wlodarczyk,2005Berberan}.
A comparison with Eq. \ref{multiint3} shows that this is not true
and reveals that $\sigma(\Gamma)$ contains information about both
the radiative and non-radiative rates:
\begin{equation}\label{multiint4}
\sigma(\Gamma_{tot})=\rho(\Gamma_{tot})\int_{0}^{\Gamma_{tot}}\rho_{\Gamma_{tot}}(\Gamma_{rad})\,\,\Gamma_{rad}d\Gamma_{rad}
\end{equation}
Thus $\sigma(\Gamma)$ is the distribution of total decay rates
weighted by the radiative rates. This conclusion demonstrates the
practical use of Eq. \ref{multiint3}: the equation allows us to
completely interpret the distribution of rates found by modelling
with Eq. \ref{multi2}. Such a complete interpretation has not been
reported before.

\subsection{Log-normal distribution of decay rates}
Distribution functions that can be used for $\sigma(\Gamma)$ are
(sums of) normal, Lorentzian, and log-normal distribution
functions. In Fig. \ref{MultiExp}(a) the luminescence decay curve
of quantum dots is successfully modelled with Eq. \ref{multi2},
with a log-normal distribution of the rate $\Gamma$
\begin{equation}\label{lognor}
\sigma(\Gamma)= A\exp\left[-(\frac{\ln \Gamma -
\ln\Gamma_{mf}}{\gamma})^2\right]
\end{equation}
where $A$ is the normalization constant, $\Gamma_{mf}$ is the most
frequent rate constant (see Fig. \ref{Distrib}). $\gamma$ is
related to the width of the distribution:
\begin{equation}\label{lognor2}
\Delta\Gamma=2\Gamma_{mf}\sinh(\gamma)
\end{equation}
where $\Delta\Gamma$ is equal to the width of the distribution at
$\frac{1}{e}$. The most frequent rate constant $\Gamma_{mf}$ and
$\gamma$ are adjustable parameters, only one extra adjustable
parameter compared to a single-exponential model. Clearly, this
model (Eq. \ref{multi2} and \ref{lognor}) describes our
non-single-exponential experimental data extremely well. The
$\chi_{r}^{2}$ was 1.17, $\Gamma_{mf}$ was 91.7 $\mu s^{-1}$
($\frac{1}{\Gamma_{mf}}=10.9$ $ns$) and the width of the
distribution $\Delta\Gamma$ was 0.57 $ns^{-1}$. In addition to
$\Gamma_{mf}$ and $\Delta\Gamma$, an average decay rate can  be
deduced from the log-normal distribution in Fig. \ref{Distrib} .
However, this average is biased since the various components are
weighted with their quantum efficiency, as shown in Eq.
\ref{multiint4}.

Modelling with a log-normal distribution of decay rates yields
direct and clear physical parameters, for instance the shape and
width of the decay rate distribution. The log-normal function is
plotted in Fig. \ref{Distrib} (curve a). The broad distribution of
rates demonstrates the strongly non-single-exponential character
of the decay curve. In Ref. \onlinecite{2006Nikolaev} we were able
to relate the width of this broad distribution to the spatial and
orientational variations of the LDOS in inverse-opal photonic
crystals.

The log-normal model was also modelled to the decay curve from
quantum dots in suspension (Fig. \ref{SinExp}). The distribution
is plotted in Fig. \ref{Distrib} (curve b). $\Gamma_{mf}$ was 25.8
$\mu s^{-1}$ ($\frac{1}{\Gamma_{mf}}=38.8$ $ns$), close to the
lifetime deduced from the single-exponential modelling of 39.0
$\pm2.8ns$. The narrow width of the distribution $\Delta\Gamma$ of
0.079 $ns^{-1}$ is in agreement with the single-exponential
character of the decay curve.

\subsection{Stretched-exponential decay}
Besides the multi-exponential models discussed in Sections C-E,
the Kohlrausch stretched-exponential decay
model\cite{1854Kohlrauscha,1980Lindsey} is widely applied to model
non-single-exponential decay curves. The fraction of excited
emitters, i.e., the reliability function, of the Kohlrausch
stretched-exponential model is equal to:
\begin{equation}
\label{str1}\frac{c(t^{'})}{c(0)}=\exp(-(\Gamma_{str}
t^{'})^{\beta})
\end{equation}
where $\beta$ is the stretch parameter, which varies between 0 and
1, and $\Gamma_{str}$ the total decay rate in case of
stretched-exponential decay. The stretch parameter $\beta$
qualitatively expresses the underlying distribution of rates: a
small $\beta$ means that the distribution of rates is broad and
$\beta$ close to 1 implies a narrow distribution. The recovery of
the distribution of rates in case of stretched-exponential decay
is mathematically complicated\footnote{In case of the
stretched-exponential model the distribution of the rates is
unknown and is generally deduced by solving the following equation
\cite{1985Jamesc,1986James,2001Lee,2002Schlegel,2005Berberan}:
$\frac{\beta}{t}(\Gamma_{str} t)^{\beta}\exp(-(\Gamma_{str}
t)^{\beta})=\int_{0}^{\infty}\sigma(\Gamma)\exp(-\Gamma t)d\Gamma$
where $\sigma(\Gamma)$ is the distribution function of total decay
rate weighted by $\Gamma_{rad}$. To deduce $\sigma(\Gamma)$ an
inverse Laplace transform is applied. For $\beta\neq0.5$ and
$\beta\neq1$ there is no analytical solution of this equation and
therefore it is difficult to deduce the distribution function.
This difficulty can be circumvented by modelling directly with a
known distribution function, as is shown in this paper.} and only
feasible for specific
$\beta$'s\cite{1980Lindsey,1985Huber,1990Siemiarczuk,1991Alvarez}.

The decay curve corresponding to a Kohlrausch
stretched-exponential decay of the fraction
$\frac{c(t^{'})}{c(0)}$ can be deduced using Eq. \ref{str1} and
Eq. \ref{master}, and results in:
\begin{equation}
\label{str2} g(t)=\frac{\beta}{t}(\Gamma_{str}
t)^{\beta}\exp(-(\Gamma_{str} t)^{\beta})
\end{equation}
The normalization of Eq. \ref{str2} can, in analogy with Eq.
\ref{norml}, be deduced by integration for all times between $t=0$
and $t\rightarrow\infty$, which yields 1. Therefore, an important
consequence is that Eq. \ref{str2} is correctly normalized
\emph{only} for emitters with a quantum yield of 1
($\Gamma_{rad}=\Gamma_{tot}$ and $f(t)=g(t)$). It is not clear how
normalization should be done in realistic cases with quantum yield
$<100\%$. To the best of our knowledge, this problem has been
overlooked in the literature.

The main advantage of modelling with a Kohlrausch
stretched-exponential function is that the average decay time
$<t>$ can readily be calculated. The average decay time is equal
to\cite{1980Lindsey}:
\begin{equation}
\label{str3} <t>=\tau_{av}=\frac{\int_{0}^\infty
g(t)tdt}{\int_{0}^\infty
g(t)dt}=\frac{1}{\Gamma_{str}\beta}\overline{\Gamma}[\frac{1}{\beta}]
\end{equation}
where $\overline{\Gamma}$ is the mathematical Gamma-function. For
the single-exponential limit of $\beta\rightarrow1$ Eq.~\ref{str3}
reduces to Eq.~\ref{sinexp2}. Note again that in this average the
various contributions are weighted with their quantum efficiency
and that the average decay time $<t>$ differs from
$\frac{1}{\Gamma{str}}$ (see Section E). Indeed, for the data in
Fig. \ref{MultiExp}(b) Eq.~\ref{str3} yielded an average decay
time of 31.1 $ns$ , strongly different from the
$\frac{1}{\Gamma_{str}}$-value of 10.4 $ns$, in contrast to the
result (Eq. \ref{sinexp1}) for single-exponential decay.

It is important to note that Eq.~\ref{str3} is the average decay
time $<t>$ corresponding to the decay curve given by
Eq.~\ref{str2}. In, for instance,
Refs.~\onlinecite{2002Schlegel,2006Kalkman} the average time given
by Eq.~\ref{str3} has erroneously been associated to fluorescence
decay described by Eq.~\ref{str1}.

In contrast to the single-exponential model, the reliability
function and the probability density function of a
stretched-exponential do \emph{not} have the same form (see Fig.
\ref{StrExpFg}); the probability density function contains a
time-dependent pre-factor. Therefore, the relation between the
reliability function and the probability density function (Eq.
\ref{master}) has important consequences. For a $\beta$-value of
0.5 the average decay time of the reliability function (Eq.
\ref{str1}) and of the probability density function (Eq.
\ref{str2}) differ by more than a factor of ten. Thus it is
important to take into consideration whether Eq. \ref{str1} or Eq.
\ref{str2} is used to describe the experimental photoluminescence
decay curve. This is important since in many reports
\cite{2001Lee,2002Lee,2002Schlegel,2003Chen,2004Fisher,2005Berberan},
the luminescence decay curve in modelled with the time dependence
of Eq. \ref{str1}. We remark that while Eq. \ref{str1} can be used
to account for the deviation from single-exponential decay, it
does not represent the true Kohlrausch function, but is simply an
alternative model. We argue that using the Kohlrausch
stretched-exponential as a reliability function to model the
fraction $\frac{c(t^{'})}{c(0)}$\footnote{We remark that in Refs.
\onlinecite{1854Kohlrauscha,1980Lindsey} capacitor discharge and
dielectric relaxation are studied, which are, in contrast to a
fluorescence decay curve, indeed described by a reliability
function.} implies that the proper probability density function,
i.e., Eq. \ref{str2}, must be used to model a luminescence decay
curve. Fig. \ref{MultiExp}(b) shows the modelling of experimental
data with Eq. \ref{str2}, with $\Gamma_{str}$ and $\beta$ as
adjustable parameters. The $\beta$-value was 0.42 and
$\Gamma_{str}$ was 96.2 $\mu s^{-1}$
($\frac{1}{\Gamma_{str}}=10.4$ $ns$). Modelling with
stretched-exponential is obviously more satisfactory than
single-exponential, but here fails at long times, reflected by the
high $\chi_{r}^2$-value of 60.7.


\section{Conclusions}
We have presented a statistical analysis of time-resolved
spontaneous emission decay curves from ensembles of emitters, in
particular colloidal quantum dots, with the aim to interpret
ubiquitous non-single-exponential decay. Contrary to what is
widely assumed, the density of excited emitters $c(t)$ and the
intensity in an emission decay curve ($f(t)$ or $g(t)$) are not
proportional, but the density is a time-integral of the intensity.
The integral relation is crucial to correctly interpret
non-single-exponential decay. We have derived the proper
normalization for both a discrete, and a continuous distribution
of rates, where every decay component is multiplied with its
radiative decay rate. A central result of our paper is the
derivation of the emission decay curve $f(t)$ in case that both
radiative and non-radiative decays are independently distributed
(Eq. \ref{multiint3}). In this case, the well-known emission
quantum efficiency can not be expressed by a single number
anymore, but it is also distributed. We derive a practical
description of non-single-exponential emission decay curves in
terms of a distribution of total decay rates weighted with the
radiative rates. Analyzing decay curves in terms of decay rate
distributions opposes to the usual and widely reported analysis in
terms of distributed lifetimes. We apply our analysis to recent
examples of colloidal quantum dot emission in suspensions and in
photonic crystals, and we find that this important class of
emitters is well described by a log-normal distribution of decay
rates with a narrow and a broad distribution, respectively.
Finally, we briefly discuss the Kohlrausch stretched-exponential
model; we deduce the average decay time and we find that its
normalization is ill-defined for emitters with a realistic quantum
efficiency of less than 100 $\%$.

\section{Acknowledgments}
This work is part of the research program of both the "Stichting
voor Fundamenteel Onderzoek der Materie (FOM)", and "Chemische
Wetenschappen", which are financially supported by the
"Nederlandse Organisatie voor Wetenschappelijk Onderzoek (NWO)".


\begin{thebibliography}{44}
\expandafter\ifx\csname
natexlab\endcsname\relax\def\natexlab#1{#1}\fi
\expandafter\ifx\csname bibnamefont\endcsname\relax
  \def\bibnamefont#1{#1}\fi
\expandafter\ifx\csname bibfnamefont\endcsname\relax
  \def\bibfnamefont#1{#1}\fi
\expandafter\ifx\csname citenamefont\endcsname\relax
  \def\citenamefont#1{#1}\fi
\expandafter\ifx\csname url\endcsname\relax
  \def\url#1{\texttt{#1}}\fi
\expandafter\ifx\csname
urlprefix\endcsname\relax\def\urlprefix{URL }\fi
\providecommand{\bibinfo}[2]{#2}
\providecommand{\eprint}[2][]{\url{#2}}

\bibitem[{\citenamefont{Lakowicz}(1999)}]{Lakowicz}
\bibinfo{author}{\bibfnamefont{J.~R.} \bibnamefont{Lakowicz}},
  \emph{\bibinfo{title}{Principles of Fluorescence Spectroscopy}}
  (\bibinfo{publisher}{Kluwer Academic/Plenum Publishers},
  \bibinfo{address}{New York, Boston, Dordrecht, London, Moscow},
  \bibinfo{year}{1999}), \bibinfo{edition}{2nd} ed.

\bibitem[{\citenamefont{Medintz et~al.}(2005)\citenamefont{Medintz, Uyeda,
  Goldman, and Mattoussi}}]{2005Medintz}
\bibinfo{author}{\bibfnamefont{I.~L.} \bibnamefont{Medintz}},
  \bibinfo{author}{\bibfnamefont{H.~T.} \bibnamefont{Uyeda}},
  \bibinfo{author}{\bibfnamefont{E.~R.} \bibnamefont{Goldman}},
  \bibnamefont{and}
  \bibinfo{author}{\bibfnamefont{H.}~\bibnamefont{Mattoussi}},
  \bibinfo{journal}{Nature Materials} \textbf{\bibinfo{volume}{4}},
  \bibinfo{pages}{435} (\bibinfo{year}{2005}).

\bibitem[{\citenamefont{Loudon}(2001)}]{Loudon}
\bibinfo{author}{\bibfnamefont{R.}~\bibnamefont{Loudon}},
  \emph{\bibinfo{title}{The quantum theory of light}}, Oxford science
  publications (\bibinfo{publisher}{Oxford University Press},
  \bibinfo{address}{Oxford,}, \bibinfo{year}{2001}), \bibinfo{edition}{3rd} ed.

\bibitem[{\citenamefont{Crooker et~al.}(2002)\citenamefont{Crooker,
  Hollingsworth, Tretiak, and Klimov}}]{2002Crooker}
\bibinfo{author}{\bibfnamefont{S.~A.} \bibnamefont{Crooker}},
  \bibinfo{author}{\bibfnamefont{J.~A.} \bibnamefont{Hollingsworth}},
  \bibinfo{author}{\bibfnamefont{S.}~\bibnamefont{Tretiak}}, \bibnamefont{and}
  \bibinfo{author}{\bibfnamefont{V.~I.} \bibnamefont{Klimov}},
  \bibinfo{journal}{Physical Review Letters} \textbf{\bibinfo{volume}{89}},
  \bibinfo{pages}{186802} (\bibinfo{year}{2002}).

\bibitem[{\citenamefont{Zhang et~al.}(2002)\citenamefont{Zhang, Wang, and
  Xiao}}]{2002Zhang}
\bibinfo{author}{\bibfnamefont{J.~Y.} \bibnamefont{Zhang}},
  \bibinfo{author}{\bibfnamefont{X.~Y.} \bibnamefont{Wang}}, \bibnamefont{and}
  \bibinfo{author}{\bibfnamefont{M.}~\bibnamefont{Xiao}},
  \bibinfo{journal}{Optics Letters} \textbf{\bibinfo{volume}{27}},
  \bibinfo{pages}{1253} (\bibinfo{year}{2002}).

\bibitem[{\citenamefont{Drexhage}(1970)}]{1970Drexhage}
\bibinfo{author}{\bibfnamefont{K.~H.} \bibnamefont{Drexhage}},
  \bibinfo{journal}{Journal of Luminescence} \textbf{\bibinfo{volume}{1,2}},
  \bibinfo{pages}{693} (\bibinfo{year}{1970}).

\bibitem[{\citenamefont{Amos and Barnes}(1997)}]{1997Amos}
\bibinfo{author}{\bibfnamefont{R.~M.} \bibnamefont{Amos}} \bibnamefont{and}
  \bibinfo{author}{\bibfnamefont{W.~L.} \bibnamefont{Barnes}},
  \bibinfo{journal}{Physical Review B} \textbf{\bibinfo{volume}{55}},
  \bibinfo{pages}{7249} (\bibinfo{year}{1997}).

\bibitem[{\citenamefont{Danz et~al.}(2002)\citenamefont{Danz, Heber, and
  Brauer}}]{2002Danz}
\bibinfo{author}{\bibfnamefont{N.}~\bibnamefont{Danz}},
  \bibinfo{author}{\bibfnamefont{J.}~\bibnamefont{Heber}}, \bibnamefont{and}
  \bibinfo{author}{\bibfnamefont{A.}~\bibnamefont{Brauer}},
  \bibinfo{journal}{Physical Review A} \textbf{\bibinfo{volume}{66}},
  \bibinfo{pages}{063809} (\bibinfo{year}{2002}).

\bibitem[{\citenamefont{Astilean and Barnes}(2002)}]{2002Astilean}
\bibinfo{author}{\bibfnamefont{S.}~\bibnamefont{Astilean}} \bibnamefont{and}
  \bibinfo{author}{\bibfnamefont{W.~L.} \bibnamefont{Barnes}},
  \bibinfo{journal}{Applied Physics B-Lasers and Optics}
  \textbf{\bibinfo{volume}{75}}, \bibinfo{pages}{591} (\bibinfo{year}{2002}).

\bibitem[{\citenamefont{Bayer et~al.}(2001)\citenamefont{Bayer, Reinecke,
  Weidner, Larionov, McDonald, and Forchel}}]{2001Bayer}
\bibinfo{author}{\bibfnamefont{M.}~\bibnamefont{Bayer}},
  \bibinfo{author}{\bibfnamefont{T.~L.} \bibnamefont{Reinecke}},
  \bibinfo{author}{\bibfnamefont{F.}~\bibnamefont{Weidner}},
  \bibinfo{author}{\bibfnamefont{A.}~\bibnamefont{Larionov}},
  \bibinfo{author}{\bibfnamefont{A.}~\bibnamefont{McDonald}}, \bibnamefont{and}
  \bibinfo{author}{\bibfnamefont{A.}~\bibnamefont{Forchel}},
  \bibinfo{journal}{Physical Review Letters} \textbf{\bibinfo{volume}{86}},
  \bibinfo{pages}{3168} (\bibinfo{year}{2001}).

\bibitem[{\citenamefont{Biteen et~al.}(2005)\citenamefont{Biteen, Pacifici,
  Lewis, and Atwater}}]{2005Biteen}
\bibinfo{author}{\bibfnamefont{J.~S.} \bibnamefont{Biteen}},
  \bibinfo{author}{\bibfnamefont{D.}~\bibnamefont{Pacifici}},
  \bibinfo{author}{\bibfnamefont{N.~S.} \bibnamefont{Lewis}}, \bibnamefont{and}
  \bibinfo{author}{\bibfnamefont{H.~A.} \bibnamefont{Atwater}},
  \bibinfo{journal}{Nano Letters} \textbf{\bibinfo{volume}{5}},
  \bibinfo{pages}{1768} (\bibinfo{year}{2005}).

\bibitem[{\citenamefont{Song et~al.}(2005)\citenamefont{Song, Atay, Shi, Urabe,
  and Nurmikko}}]{2005Song}
\bibinfo{author}{\bibfnamefont{J.~H.} \bibnamefont{Song}},
  \bibinfo{author}{\bibfnamefont{T.}~\bibnamefont{Atay}},
  \bibinfo{author}{\bibfnamefont{S.~F.} \bibnamefont{Shi}},
  \bibinfo{author}{\bibfnamefont{H.}~\bibnamefont{Urabe}}, \bibnamefont{and}
  \bibinfo{author}{\bibfnamefont{A.~V.} \bibnamefont{Nurmikko}},
  \bibinfo{journal}{Nano Letters} \textbf{\bibinfo{volume}{5}},
  \bibinfo{pages}{1557} (\bibinfo{year}{2005}).

\bibitem[{\citenamefont{Fujita et~al.}(2005)\citenamefont{Fujita, Takahashi,
  Tanaka, Asano, and Noda}}]{2005Fujita}
\bibinfo{author}{\bibfnamefont{M.}~\bibnamefont{Fujita}},
  \bibinfo{author}{\bibfnamefont{S.}~\bibnamefont{Takahashi}},
  \bibinfo{author}{\bibfnamefont{Y.}~\bibnamefont{Tanaka}},
  \bibinfo{author}{\bibfnamefont{T.}~\bibnamefont{Asano}}, \bibnamefont{and}
  \bibinfo{author}{\bibfnamefont{S.}~\bibnamefont{Noda}},
  \bibinfo{journal}{Science} \textbf{\bibinfo{volume}{308}},
  \bibinfo{pages}{1296} (\bibinfo{year}{2005}).

\bibitem[{\citenamefont{Kress et~al.}(2005)\citenamefont{Kress, Hofbauer,
  Reinelt, Kaniber, Krenner, Meyer, Bohm, and Finley}}]{2005Kress}
\bibinfo{author}{\bibfnamefont{A.}~\bibnamefont{Kress}},
  \bibinfo{author}{\bibfnamefont{F.}~\bibnamefont{Hofbauer}},
  \bibinfo{author}{\bibfnamefont{N.}~\bibnamefont{Reinelt}},
  \bibinfo{author}{\bibfnamefont{M.}~\bibnamefont{Kaniber}},
  \bibinfo{author}{\bibfnamefont{H.~J.} \bibnamefont{Krenner}},
  \bibinfo{author}{\bibfnamefont{R.}~\bibnamefont{Meyer}},
  \bibinfo{author}{\bibfnamefont{G.}~\bibnamefont{Bohm}}, \bibnamefont{and}
  \bibinfo{author}{\bibfnamefont{J.~J.} \bibnamefont{Finley}},
  \bibinfo{journal}{Physical Review B} \textbf{\bibinfo{volume}{71}},
  \bibinfo{pages}{241304} (\bibinfo{year}{2005}).

\bibitem[{\citenamefont{Lodahl et~al.}(2004)\citenamefont{Lodahl, van Driel,
  Nikolaev, Irman, Overgaag, Vanmaekelbergh, and Vos}}]{2004Lodahl}
\bibinfo{author}{\bibfnamefont{P.}~\bibnamefont{Lodahl}},
  \bibinfo{author}{\bibfnamefont{A.~F.} \bibnamefont{van Driel}},
  \bibinfo{author}{\bibfnamefont{I.~S.} \bibnamefont{Nikolaev}},
  \bibinfo{author}{\bibfnamefont{A.}~\bibnamefont{Irman}},
  \bibinfo{author}{\bibfnamefont{K.}~\bibnamefont{Overgaag}},
  \bibinfo{author}{\bibfnamefont{D.}~\bibnamefont{Vanmaekelbergh}},
  \bibnamefont{and} \bibinfo{author}{\bibfnamefont{W.~L.} \bibnamefont{Vos}},
  \bibinfo{journal}{Nature} \textbf{\bibinfo{volume}{430}},
  \bibinfo{pages}{654} (\bibinfo{year}{2004}).

\bibitem[{\citenamefont{Foggi et~al.}(1995)\citenamefont{Foggi, Pettini, Santa,
  Righini, and Califano}}]{1995Foggi}
\bibinfo{author}{\bibfnamefont{P.}~\bibnamefont{Foggi}},
  \bibinfo{author}{\bibfnamefont{L.}~\bibnamefont{Pettini}},
  \bibinfo{author}{\bibfnamefont{I.}~\bibnamefont{Santa}},
  \bibinfo{author}{\bibfnamefont{R.}~\bibnamefont{Righini}}, \bibnamefont{and}
  \bibinfo{author}{\bibfnamefont{S.}~\bibnamefont{Califano}},
  \bibinfo{journal}{Journal of Physical Chemistry}
  \textbf{\bibinfo{volume}{99}}, \bibinfo{pages}{7439} (\bibinfo{year}{1995}).

\bibitem[{\citenamefont{Klimov and McBranch}(1998)}]{1998Klimov}
\bibinfo{author}{\bibfnamefont{V.~I.} \bibnamefont{Klimov}} \bibnamefont{and}
  \bibinfo{author}{\bibfnamefont{D.~W.} \bibnamefont{McBranch}},
  \bibinfo{journal}{Physical Review Letters} \textbf{\bibinfo{volume}{80}},
  \bibinfo{pages}{4028} (\bibinfo{year}{1998}).

\bibitem[{\citenamefont{Neuwahl et~al.}(2000)\citenamefont{Neuwahl, Foggi, and
  Brown}}]{2000Neuwahl}
\bibinfo{author}{\bibfnamefont{F.~V.~R.} \bibnamefont{Neuwahl}},
  \bibinfo{author}{\bibfnamefont{P.}~\bibnamefont{Foggi}}, \bibnamefont{and}
  \bibinfo{author}{\bibfnamefont{R.~G.} \bibnamefont{Brown}},
  \bibinfo{journal}{Chemical Physics Letters} \textbf{\bibinfo{volume}{319}},
  \bibinfo{pages}{157} (\bibinfo{year}{2000}).

\bibitem[{\citenamefont{Rosencwaig}(1980)}]{Rosencwaig}
\bibinfo{author}{\bibfnamefont{A.}~\bibnamefont{Rosencwaig}},
  \emph{\bibinfo{title}{Photoacoustics and Photoacoustic Spectroscopy}}
  (\bibinfo{publisher}{John Wiley \& Sons}, \bibinfo{address}{New York},
  \bibinfo{year}{1980}).

\bibitem[{\citenamefont{Grinberg et~al.}(2003)\citenamefont{Grinberg, Sikorska,
  and Sliwinski}}]{2003Grinberg}
\bibinfo{author}{\bibfnamefont{M.}~\bibnamefont{Grinberg}},
  \bibinfo{author}{\bibfnamefont{A.}~\bibnamefont{Sikorska}}, \bibnamefont{and}
  \bibinfo{author}{\bibfnamefont{A.}~\bibnamefont{Sliwinski}},
  \bibinfo{journal}{Physical Review B} \textbf{\bibinfo{volume}{67}},
  \bibinfo{pages}{045114} (\bibinfo{year}{2003}).

\bibitem[{\citenamefont{James and Ware}(1986)}]{1986James}
\bibinfo{author}{\bibfnamefont{D.~R.} \bibnamefont{James}} \bibnamefont{and}
  \bibinfo{author}{\bibfnamefont{W.~R.} \bibnamefont{Ware}},
  \bibinfo{journal}{Chemical Physics Letters} \textbf{\bibinfo{volume}{126}},
  \bibinfo{pages}{7} (\bibinfo{year}{1986}).

\bibitem[{\citenamefont{Siemiarczuk et~al.}(1990)\citenamefont{Siemiarczuk,
  Wagner, and Ware}}]{1990Siemiarczuk}
\bibinfo{author}{\bibfnamefont{A.}~\bibnamefont{Siemiarczuk}},
  \bibinfo{author}{\bibfnamefont{B.~D.} \bibnamefont{Wagner}},
  \bibnamefont{and} \bibinfo{author}{\bibfnamefont{W.~R.} \bibnamefont{Ware}},
  \bibinfo{journal}{Journal of Physical Chemistry}
  \textbf{\bibinfo{volume}{94}}, \bibinfo{pages}{1661} (\bibinfo{year}{1990}).

\bibitem[{\citenamefont{Brochon et~al.}(1990)\citenamefont{Brochon, Livesey,
  Pouget, and Valeur}}]{1990Brochon}
\bibinfo{author}{\bibfnamefont{J.~C.} \bibnamefont{Brochon}},
  \bibinfo{author}{\bibfnamefont{A.~K.} \bibnamefont{Livesey}},
  \bibinfo{author}{\bibfnamefont{J.}~\bibnamefont{Pouget}}, \bibnamefont{and}
  \bibinfo{author}{\bibfnamefont{B.}~\bibnamefont{Valeur}},
  \bibinfo{journal}{Chemical Physics Letters} \textbf{\bibinfo{volume}{174}},
  \bibinfo{pages}{517} (\bibinfo{year}{1990}).

\bibitem[{\citenamefont{Wlodarczyk and Kierdaszuk}(2003)}]{2003Wlodarczyk}
\bibinfo{author}{\bibfnamefont{J.}~\bibnamefont{Wlodarczyk}} \bibnamefont{and}
  \bibinfo{author}{\bibfnamefont{B.}~\bibnamefont{Kierdaszuk}},
  \bibinfo{journal}{Biophysical Journal} \textbf{\bibinfo{volume}{85}},
  \bibinfo{pages}{589} (\bibinfo{year}{2003}).

\bibitem[{\citenamefont{Wuister et~al.}(2004)\citenamefont{Wuister, van
  Houselt, Donega, Vanmaekelbergh, and Meijerink}}]{2004Wuister}
\bibinfo{author}{\bibfnamefont{S.~F.} \bibnamefont{Wuister}},
  \bibinfo{author}{\bibfnamefont{A.}~\bibnamefont{van Houselt}},
  \bibinfo{author}{\bibfnamefont{C.~D.~M.} \bibnamefont{Donega}},
  \bibinfo{author}{\bibfnamefont{D.}~\bibnamefont{Vanmaekelbergh}},
  \bibnamefont{and}
  \bibinfo{author}{\bibfnamefont{A.}~\bibnamefont{Meijerink}},
  \bibinfo{journal}{Angewandte Chemie-International Edition}
  \textbf{\bibinfo{volume}{43}}, \bibinfo{pages}{3029} (\bibinfo{year}{2004}).

\bibitem[{\citenamefont{Nikolaev et~al.}(2005)\citenamefont{Nikolaev, Lodahl,
  van Driel, and Vos}}]{2006Nikolaev}
\bibinfo{author}{\bibfnamefont{I.~S.} \bibnamefont{Nikolaev}},
  \bibinfo{author}{\bibfnamefont{P.}~\bibnamefont{Lodahl}},
  \bibinfo{author}{\bibfnamefont{A.~F.} \bibnamefont{van Driel}},
  \bibnamefont{and} \bibinfo{author}{\bibfnamefont{W.~L.} \bibnamefont{Vos}},
  \bibinfo{journal}{http://arxiv.org/abs/physics/0511133}
  (\bibinfo{year}{2005}).

\bibitem[{\citenamefont{Kohlrausch}(1854{\natexlab{a}})}]{1854Kohlrauscha}
\bibinfo{author}{\bibfnamefont{R.}~\bibnamefont{Kohlrausch}},
  \bibinfo{journal}{Annalen der Physik} \textbf{\bibinfo{volume}{91}},
  \bibinfo{pages}{179} (\bibinfo{year}{1854}{\natexlab{a}}).

\bibitem[{\citenamefont{Deschenes and Vanden~Bout}(2001)}]{2001Deschenes}
\bibinfo{author}{\bibfnamefont{L.~A.} \bibnamefont{Deschenes}}
  \bibnamefont{and} \bibinfo{author}{\bibfnamefont{D.~A.}
  \bibnamefont{Vanden~Bout}}, \bibinfo{journal}{Science}
  \textbf{\bibinfo{volume}{292}}, \bibinfo{pages}{255} (\bibinfo{year}{2001}).

\bibitem[{\citenamefont{Lindsey and Patterson}(1980)}]{1980Lindsey}
\bibinfo{author}{\bibfnamefont{C.~P.} \bibnamefont{Lindsey}} \bibnamefont{and}
  \bibinfo{author}{\bibfnamefont{G.~D.} \bibnamefont{Patterson}},
  \bibinfo{journal}{Journal of Chemical Physics} \textbf{\bibinfo{volume}{73}},
  \bibinfo{pages}{3348} (\bibinfo{year}{1980}).

\bibitem[{\citenamefont{Kohlrausch}(1854{\natexlab{b}})}]{1854Kohlrauschb}
\bibinfo{author}{\bibfnamefont{R.}~\bibnamefont{Kohlrausch}},
  \bibinfo{journal}{Annalen der Physik} \textbf{\bibinfo{volume}{91}},
  \bibinfo{pages}{56} (\bibinfo{year}{1854}{\natexlab{b}}).

\bibitem[{\citenamefont{Torre et~al.}(2004)\citenamefont{Torre, Bartolini, and
  Righini}}]{2004Torre}
\bibinfo{author}{\bibfnamefont{R.}~\bibnamefont{Torre}},
  \bibinfo{author}{\bibfnamefont{P.}~\bibnamefont{Bartolini}},
  \bibnamefont{and} \bibinfo{author}{\bibfnamefont{R.}~\bibnamefont{Righini}},
  \bibinfo{journal}{Nature} \textbf{\bibinfo{volume}{428}},
  \bibinfo{pages}{296} (\bibinfo{year}{2004}).

\bibitem[{\citenamefont{Schlegel et~al.}(2002)\citenamefont{Schlegel,
  Bohnenberger, Potapova, and Mews}}]{2002Schlegel}
\bibinfo{author}{\bibfnamefont{G.}~\bibnamefont{Schlegel}},
  \bibinfo{author}{\bibfnamefont{J.}~\bibnamefont{Bohnenberger}},
  \bibinfo{author}{\bibfnamefont{I.}~\bibnamefont{Potapova}}, \bibnamefont{and}
  \bibinfo{author}{\bibfnamefont{A.}~\bibnamefont{Mews}},
  \bibinfo{journal}{Physical Review Letters} \textbf{\bibinfo{volume}{88}},
  \bibinfo{pages}{137401} (\bibinfo{year}{2002}).

\bibitem[{\citenamefont{Chen}(2003)}]{2003Chen}
\bibinfo{author}{\bibfnamefont{R.}~\bibnamefont{Chen}},
  \bibinfo{journal}{Journal of Luminescence} \textbf{\bibinfo{volume}{102}},
  \bibinfo{pages}{510} (\bibinfo{year}{2003}).

\bibitem[{\citenamefont{Fisher et~al.}(2004)\citenamefont{Fisher, Eisler,
  Stott, and Bawendi}}]{2004Fisher}
\bibinfo{author}{\bibfnamefont{B.~R.} \bibnamefont{Fisher}},
  \bibinfo{author}{\bibfnamefont{H.~J.} \bibnamefont{Eisler}},
  \bibinfo{author}{\bibfnamefont{N.~E.} \bibnamefont{Stott}}, \bibnamefont{and}
  \bibinfo{author}{\bibfnamefont{M.~G.} \bibnamefont{Bawendi}},
  \bibinfo{journal}{Journal of Physical Chemistry B}
  \textbf{\bibinfo{volume}{108}}, \bibinfo{pages}{143} (\bibinfo{year}{2004}).

\bibitem[{\citenamefont{Huber}(1985)}]{1985Huber}
\bibinfo{author}{\bibfnamefont{D.~L.} \bibnamefont{Huber}},
  \bibinfo{journal}{Physical Review B} \textbf{\bibinfo{volume}{31}},
  \bibinfo{pages}{6070} (\bibinfo{year}{1985}).

\bibitem[{\citenamefont{Alvarez et~al.}(1991)\citenamefont{Alvarez, Alegria,
  and Colmenero}}]{1991Alvarez}
\bibinfo{author}{\bibfnamefont{F.}~\bibnamefont{Alvarez}},
  \bibinfo{author}{\bibfnamefont{A.}~\bibnamefont{Alegria}}, \bibnamefont{and}
  \bibinfo{author}{\bibfnamefont{J.}~\bibnamefont{Colmenero}},
  \bibinfo{journal}{Physical Review B} \textbf{\bibinfo{volume}{44}},
  \bibinfo{pages}{7306} (\bibinfo{year}{1991}).

\bibitem[{\citenamefont{Lee et~al.}(2002)\citenamefont{Lee, Kim, Tang, and
  Hochstrasser}}]{2002Lee}
\bibinfo{author}{\bibfnamefont{M.}~\bibnamefont{Lee}},
  \bibinfo{author}{\bibfnamefont{J.}~\bibnamefont{Kim}},
  \bibinfo{author}{\bibfnamefont{J.}~\bibnamefont{Tang}}, \bibnamefont{and}
  \bibinfo{author}{\bibfnamefont{R.~M.} \bibnamefont{Hochstrasser}},
  \bibinfo{journal}{Chemical Physics Letters} \textbf{\bibinfo{volume}{359}},
  \bibinfo{pages}{412} (\bibinfo{year}{2002}).

\bibitem[{\citenamefont{Vergeer et~al.}(2005)\citenamefont{Vergeer, Vlugt, Kox,
  den Hertog, van~der Eerden, and Meijerink}}]{2005Vergeer}
\bibinfo{author}{\bibfnamefont{P.}~\bibnamefont{Vergeer}},
  \bibinfo{author}{\bibfnamefont{T.~J.~H.} \bibnamefont{Vlugt}},
  \bibinfo{author}{\bibfnamefont{M.~H.~F.} \bibnamefont{Kox}},
  \bibinfo{author}{\bibfnamefont{M.~I.} \bibnamefont{den Hertog}},
  \bibinfo{author}{\bibfnamefont{J.}~\bibnamefont{van~der Eerden}},
  \bibnamefont{and}
  \bibinfo{author}{\bibfnamefont{A.}~\bibnamefont{Meijerink}},
  \bibinfo{journal}{Physical Review B} \textbf{\bibinfo{volume}{71}},
  \bibinfo{pages}{014119} (\bibinfo{year}{2005}).

\bibitem[{\citenamefont{Dougherty}(1990)}]{Dougherty}
\bibinfo{author}{\bibfnamefont{E.~R.} \bibnamefont{Dougherty}},
  \emph{\bibinfo{title}{Probability and Statistics for the engineering,
  computing and physical sciences}} (\bibinfo{publisher}{Prentice-Hall
  International, Inc.}, \bibinfo{address}{Englewood, New Jersey},
  \bibinfo{year}{1990}).

\bibitem[{\citenamefont{van Driel et~al.}(2005)\citenamefont{van Driel, Allan,
  Delerue, Lodahl, Vos, and Vanmaekelbergh}}]{2005vanDriel}
\bibinfo{author}{\bibfnamefont{A.~F.} \bibnamefont{van Driel}},
  \bibinfo{author}{\bibfnamefont{G.}~\bibnamefont{Allan}},
  \bibinfo{author}{\bibfnamefont{C.}~\bibnamefont{Delerue}},
  \bibinfo{author}{\bibfnamefont{P.}~\bibnamefont{Lodahl}},
  \bibinfo{author}{\bibfnamefont{W.~L.} \bibnamefont{Vos}}, \bibnamefont{and}
  \bibinfo{author}{\bibfnamefont{D.}~\bibnamefont{Vanmaekelbergh}},
  \bibinfo{journal}{Physical Review Letters} \textbf{\bibinfo{volume}{95}},
  \bibinfo{pages}{236804} (\bibinfo{year}{2005}).

\bibitem[{\citenamefont{Sprik et~al.}(1996)\citenamefont{Sprik, van Tiggelen,
  and Lagendijk}}]{1996Sprik}
\bibinfo{author}{\bibfnamefont{R.}~\bibnamefont{Sprik}},
  \bibinfo{author}{\bibfnamefont{B.~A.} \bibnamefont{van Tiggelen}},
  \bibnamefont{and}
  \bibinfo{author}{\bibfnamefont{A.}~\bibnamefont{Lagendijk}},
  \bibinfo{journal}{Europhysics Letters} \textbf{\bibinfo{volume}{35}},
  \bibinfo{pages}{265} (\bibinfo{year}{1996}).

\bibitem[{\citenamefont{James et~al.}(1985)\citenamefont{James, Liu, De~Mayo,
  and Ware}}]{1985Jamesc}
\bibinfo{author}{\bibfnamefont{D.~R.} \bibnamefont{James}},
  \bibinfo{author}{\bibfnamefont{Y.-S.} \bibnamefont{Liu}},
  \bibinfo{author}{\bibfnamefont{P.}~\bibnamefont{De~Mayo}}, \bibnamefont{and}
  \bibinfo{author}{\bibfnamefont{W.~R.} \bibnamefont{Ware}},
  \bibinfo{journal}{Chemical Physics Letters} \textbf{\bibinfo{volume}{120}},
  \bibinfo{pages}{460} (\bibinfo{year}{1985}).

\bibitem[{\citenamefont{Benny~Lee et~al.}(2001)\citenamefont{Benny~Lee, Siegel,
  Webb, Leveque-Fort, Cole, Jones, Dowling, Lever, and French}}]{2001Lee}
\bibinfo{author}{\bibfnamefont{K.~C.} \bibnamefont{Benny~Lee}},
  \bibinfo{author}{\bibfnamefont{J.}~\bibnamefont{Siegel}},
  \bibinfo{author}{\bibfnamefont{S.~E.~D.} \bibnamefont{Webb}},
  \bibinfo{author}{\bibfnamefont{S.}~\bibnamefont{Leveque-Fort}},
  \bibinfo{author}{\bibfnamefont{M.~J.} \bibnamefont{Cole}},
  \bibinfo{author}{\bibfnamefont{R.}~\bibnamefont{Jones}},
  \bibinfo{author}{\bibfnamefont{K.}~\bibnamefont{Dowling}},
  \bibinfo{author}{\bibfnamefont{M.~J.} \bibnamefont{Lever}}, \bibnamefont{and}
  \bibinfo{author}{\bibfnamefont{P.~M.~W.} \bibnamefont{French}},
  \bibinfo{journal}{Biophysical Journal} \textbf{\bibinfo{volume}{81}},
  \bibinfo{pages}{1265} (\bibinfo{year}{2001}).

\bibitem[{\citenamefont{Berberan-Santos
  et~al.}(2005)\citenamefont{Berberan-Santos, Bodunov, and
  Valeur}}]{2005Berberan}
\bibinfo{author}{\bibfnamefont{M.~N.} \bibnamefont{Berberan-Santos}},
  \bibinfo{author}{\bibfnamefont{E.~N.} \bibnamefont{Bodunov}},
  \bibnamefont{and} \bibinfo{author}{\bibfnamefont{B.}~\bibnamefont{Valeur}},
  \bibinfo{journal}{Chemical Physics} \textbf{\bibinfo{volume}{315}},
  \bibinfo{pages}{171} (\bibinfo{year}{2005}).

\bibitem[{\citenamefont{Kalkman et~al.}(2006)\citenamefont{Kalkman,
  Gersen, Kuipers, and Polman}}]{2006Kalkman}
\bibinfo{author}{\bibfnamefont{J.}~\bibnamefont{Kalkman}},
  \bibinfo{author}{\bibfnamefont{H.}~\bibnamefont{Gersen}},
  \bibinfo{author}{\bibfnamefont{L.}~\bibnamefont{Kuipers}}, \bibnamefont{and}
  \bibinfo{author}{\bibfnamefont{A.}~\bibnamefont{Polman}},
  \bibinfo{journal}{Physical Review B} \textbf{\bibinfo{volume}{73}},
  \bibinfo{pages}{075317} (\bibinfo{year}{2006}).

\end{thebibliography}
\end{document}